\documentclass[aps,prb,superscriptaddress,twocolumn]{revtex4-2}
\usepackage{epigraph}
\usepackage[cp1251]{inputenc}
\usepackage[english]{babel}
\usepackage{amsmath}
\usepackage{amssymb}
\usepackage{amsfonts}
\usepackage{color}
\usepackage{graphicx}
\usepackage{lipsum,graphicx}
\usepackage{amsfonts}
\usepackage{textcomp}
\newcommand{\beq}{\begin{equation}}
	\newcommand{\eeq}{\end{equation}}
\newcommand{\bea}{\begin{eqnarray}}
	\newcommand{\eea}{\end{eqnarray}}
\usepackage{hyperref}  
\usepackage{hyperref}

\begin{document}
\hfill 
	
\vspace{20pt}		
\title{Progress, problems and prospects of room-temperature superconductivity}
\author{Ivan A. Troyan}
\affiliation{A.~V.~Shubnikov Institute of Crystallography
	of the Kurchatov Complex of Crystallography and Photonics, NRC ``Kurchatov Institute'',
	Moscow 119333, Russia}
\author{Dmitrii V. Semenok}
\affiliation{Center for High Pressure Science and Technology Advanced Research,  Beijing, 100193, China}
\author{Andrey V. Sadakov}
	\affiliation{V.~L. =~Ginzburg Research Center for High-Temperature Superconductivity and Quantum
		Materials, P. N. Lebedev Physical Institute, Russian Academy of Sciences, Moscow 119333, Russia}
	\author{Igor S. Lyubutin}
	\affiliation{A.~V.~Shubnikov Institute of Crystallography
		of the Kurchatov Complex of Crystallography and Photonics, NRC ``Kurchatov Institute'',
		Moscow 119333, Russia}
\author{Vladimir M. Pudalov}
 	\affiliation{V.~L.~Ginzburg Research Center for High-Temperature Superconductivity and Quantum
 		Materials, P.~N.~Lebedev Physical Institute, Russian Academy of Sciences, Moscow 119333, Russia}
	\affiliation{National Research Unibversity HSE, Moscow 101000, Russia}

\begin{abstract}
Discovery of superconductivity at megabar (MB) pressures in hydrogen sulfide H$_3$S, then in metal polyhydrides, starting with binary, LaH$_{10}$, etc., and ending with ternary ones, including (La,\,Y)H$_{10}$, revolutionized the field of condensed matter physics. 
These discoveries strengthen hopes  for solution of the
century-old problem of creating materials that are superconducting at room temperature. 
In experiments performed over the past 5 years at MB pressures, in addition to the synthesis of hydrides itself, their physical properties were studied using  optical, X-ray and M\"{o}ssbauer spectroscopy, as well as galvanomagnetic measurement techniques. 
This paper presents the major results of galvanomagnetic studies, including measurements in high static (up to 21\,T) and pulsed (up to 70\,T) magnetic fields.
Measurements of resistance drops to vanishingly small level at temperatures below the critical $T_c$ value,
a decrease in the critical temperature $T_c$ with increasing magnetic field, as well as diamagnetic screening, indicate the superconducting state of the polyhydrides.
The results of measurements of the isotope effect, together with the effect of magnetic impurities on $T_c$, indicate the electron-phonon mechanism of electron pairing.
However, electron-electron correlations in polyhydrides are by no means small, both in the superconducting and normal states.
It is possible that this is precisely what accounts for the unusual properties of polyhydrides that have not yet received a satisfactory explanation, such as a linear temperature dependence of the second critical field $H_{c2}(T)$, a linear dependence of resistance $\rho(T)$, 
and a linear magnetoresistance, very similar to that discovered by P.~L. Kapitza in 1929.
\end{abstract}
\maketitle
{\em This paper is devoted to the memory of P.~L.~Kapitza whose experiments in the 1920s - 1930s inspired research in Russia in high magnetic fields and at low temperatures. One of the authors (V.M.P.) is also 
grateful to P.~L.~Kapitza for the opportunity to work for several years in  the  unforgettable  creative atmosphere of the  Institute for Physical Problems.}

\section{Historical introduction. Brief chronology of 	superconductor discoveries }
The history of superconductivity (SC) started with liquefaction in 1908 by Kamerlingh Onnes
helium and subsequent (1911) discovery of the disappearance of resistance in a mercury wire immersed in liquid helium.

In the next fifty years, until the 1960s, many superconducting metals and intermetallic  compounds were discovered. The most widely used intermetallic compounds NbTi ($T_c=9.8$K) and Nb$_3$Sn ($T_c = 18$K) are representatives of type 2 superconductors - a class discovered by L.~V. Shubnikov in the 1930s. Finally, in 1986 Karl M\"{u}ller and Georg Bednorz discovered superconductivity in copper oxide ceramic compounds. In this class of compounds, the record high $T_c= 138$\,K belongs to the compound HgBaCaCuO(F), while YBa$_2$Cu$_3$O$_{7-x}$, and GdBa$_2$Cu$_3$O$_{7-x}$ with a critical temperature of $\sim 93$\,K are now widely used for practical applications.

 The most common and widespread mechanism of the Cooper pairing of electrons due to electron-phonon interaction obviously laeds to the dependense of the critical temperature $T_c$  on the mass of atoms of the crystal lattice. Since the lightest element is hydrogen, the attention of researchers has long been focused on it.
 The possibility of transformation of highly compressed hydrogen  to metal was first  conjectured
 in 1935 by E.~Wigner and  H.~B.~Huntington \cite{Wigner-Huntington_1935}. In 1968, N.~W~ Ashcroft \cite{Ashcroft_1968}, and   then, in 1989 T.~W.~Barbee et al.  \cite{Barbee_1989} predicted that the 
metallic phase of the compressed hydrogen would become supercomnding and the critical temperature of the superconducting state could reach  $\sim 200 - 400$\,K.  Atomic metallic hydrogen in solid form has not yet been relaibly obtained in static high pressure experiments, since gigantic pressures of the order of 400-500\,GPa are required for this purpose.

In 2004, Ashcroft proposed that hydrogen-rich compounds could have high critical temperatures \cite{ashcroft_PRL_2004}, and the pressures required for this should be significantly lower than the pressures required to convert purely hydrogen into a metallic superconducting state. Two years later,  J. Feng et al. \cite{feng_PRL_2006} predicted 
high-temperature superconductivity in silane (SiH$_4$). This prediction was only partially confirmed: silane indeed demonstrated a superconducting state at a pressure of 100\,GPa, but its critical temperature turned out to be only 17\,K\cite{eremets_Sci_2008}.
Nevertheless, Ashcroft's proposal stimulated an intensive experimental search for superconducting hydrides, which culminated in 2015 with the discovery of superconductivity in H$_3$S by the group of M.~E.~Eremets \cite{drozdov_Nat_2015, eremets_UFN_2016}. An increase in the critical temperature ($T_c=205$\,K), by more than 60\,K compared to that achieved in cuprate oxides, demonstrated the potential capabilities of superconducting hydrides and gave a powerful impetus to their further research.
  
  Today, many metal hydrides have already been discovered, that become superconducting at high pressure and have  critical temperatures up to $250 - 260$\,K. As a result, the superconductivity of polyhydrides has emerged and  matured as a separate and interesting field of research.
  
 \section{Key results of polyhydride research}
  \subsection{Vanishing the electrical resistance}
\begin{figure}[tbp]
  \includegraphics[width=240pt]{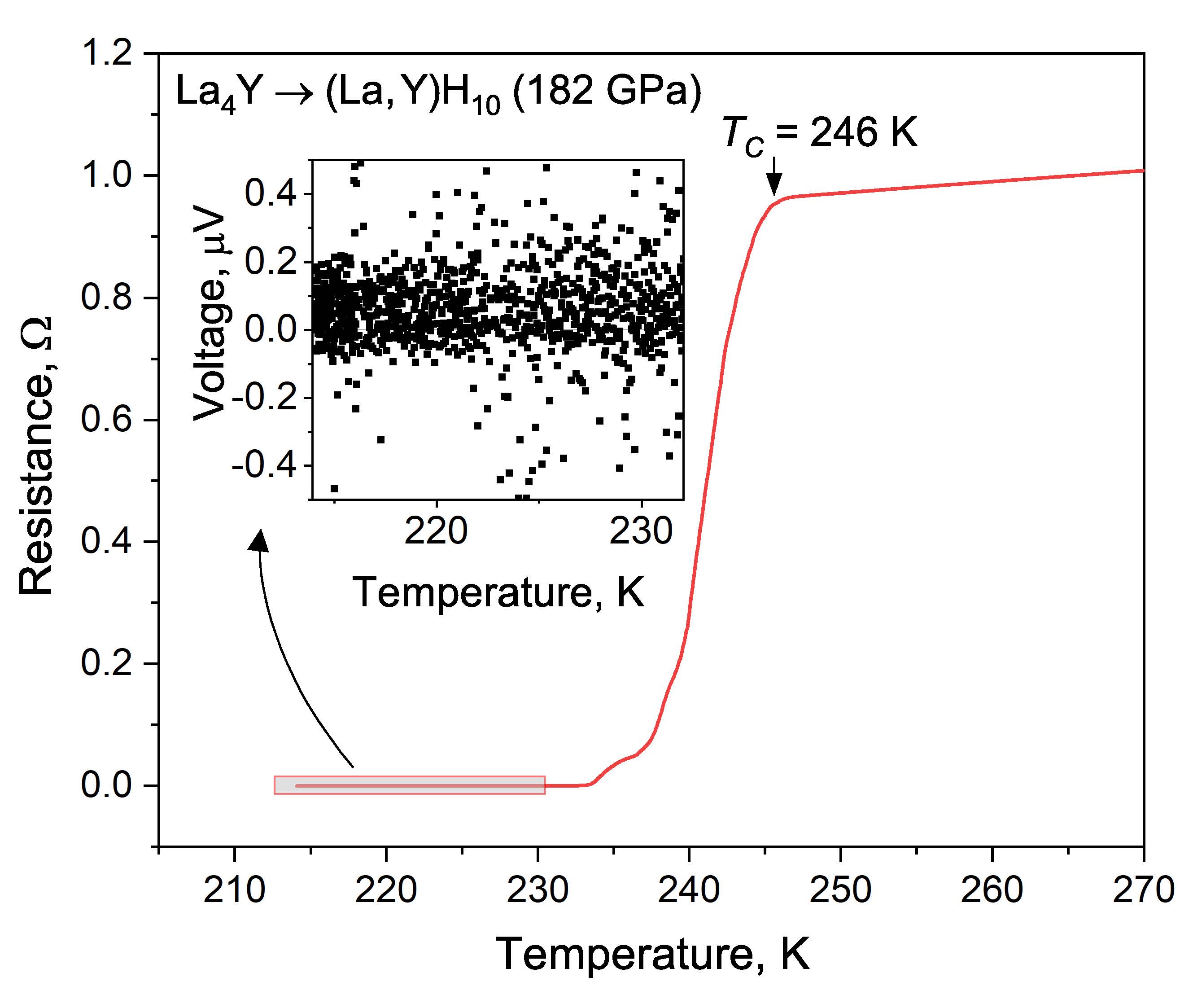}  
 \caption{Temperature dependence of resistance at the superconducting transition in (La,\,Y)H$_{10}$ sample at a pressure of 182\,GPa. The inset shows, on an enlarged scale, voltage drop between the potential psobes measured in the superconducting state with bias current of 1\,mA.
 }
  	\label{fig:R(T)_LaY}
  \end{figure}

  In hydrides of lanthanum, yttrium, thorium, etc., a sharp drop in electrical resistance is observed when  temperature decreases below a critical $T_c$, while the $T_c$ value  depends on pressure. When the 4-probe measurement technique is used, the measured resistance for $T< T_c$ is    at the level of noise, $< 0.1$m$\Omega$ \cite{semenok_MatToday_2021, troyan_UFN_2022} (see Fig.~\ref{fig:R(T)_LaY}).

 The application of an external magnetic field reduces the superconducting transition temperature and also broadens the transition itself. Due to the extremely high values of the upper critical magnetic field (which destroys superconductivity), the broadening becomes noticeable only at high fields. 
Figure \ref{fig:R(T,H)_LaY} demonstrates the effect of a magnetic field on the superconducting transition of (La,\,Y)H$_{10}$ \cite{semenok_MatToday_2021}.
  
   \begin{figure}[tbp]
		\includegraphics[width=230pt]{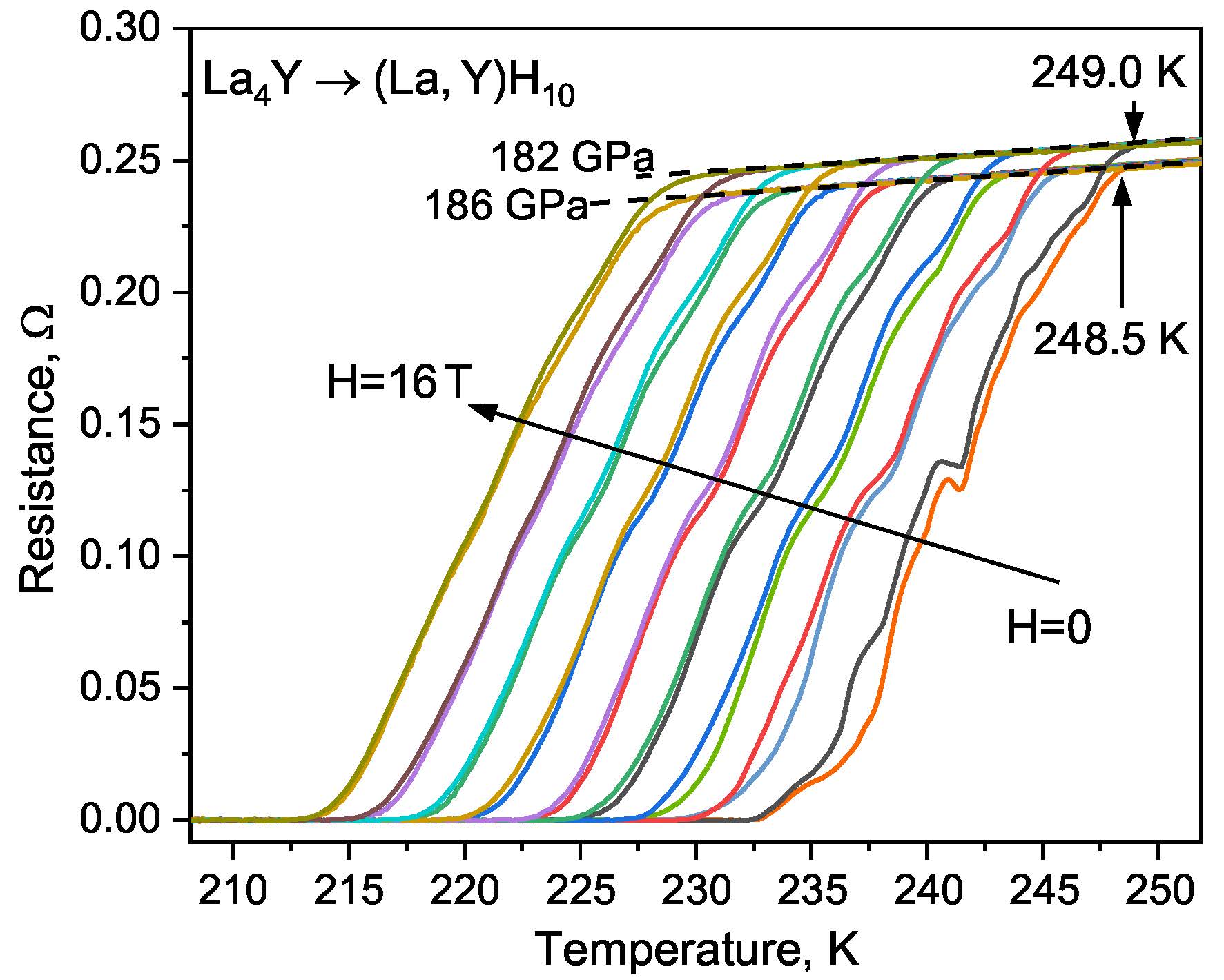} 
  	\caption{ Effect of an external magnetic field (below a critical value) on the temperature dependence of resistance	for (La,\,Y)H$_{10}$.  
  			The results are shown for two pressure values, 182 and 186\,GPa. 
  			The external magnetic field changes	(along the arrow) from 0 to 16\,T in steps of 2\,T. Numbers with vertical arrows at the top of the curves mark the critical temperatures  for two pressure values at $H=0$.
  	Adapted from \cite{semenok_MatToday_2021}.
    	}
  	\label{fig:R(T,H)_LaY}
  	 \end{figure}
 
 \subsection{Isotope-effect}
One of the most important results indicating the electron-phonon mechanism of superconductivity in hydrides is the isotope effect. This effect manifests itself in a decrease of the superconducting transition temperature when hydrogen is replaced by heavier deuterium atoms in the structure of the compound.
This effect was observed for H$_3$S \cite{drozdov_Nat_2015}, LaH$_{10}$ \cite{drozdov_Nat_2019}, YH$_6$ \cite{troyan_AdvMat_2021}, YH$_9$ \cite{kong_NatCom_2021}, CeH$_{9-10}$ \cite{chen_PRL_2021}, and a number of other compounds. In all cases, the isotope coefficient $\alpha = - \ln(T_c)/\ln(M)$, where $M$  is the  mass of atom, is within the interval from -0.3 to –0.6, in a reasonable agreement  with Fr\"{o}hlich prediction  and with the BCS theory ($\alpha=-0.5$).

A certain difficalty in the analysis is introduced by the fact that the 
ionic radius and bond energy of deuterides do not coincide with those of 
hydrides, and the limits of stability on the pressure scale and the region of distortion of the structures of hydrides and deuterides do not coincide to an even greater extent. For this reason, comparison of $T_c$ values for hydrides and deuterides at the same pressure is sometimes inaccurate, 

In general, deuterides exhibit the same properties as hydrides, namely, the superconducting transition is shifted down by the applied magnetic field. The upper critical field $H_{c2}(0)$, that is proportional to $T_c$, in deuterides, as a rule, is significantly less than in hydrides. Finally, there is a critical current, the value of which also depends on  magnetic field. When the pressure lowers,
the critical temperature of the SC transition in deuterides noticeably decreases, and then the compound decomposes with the formation of lower deuterides and D$_2$ \cite{troyan_UFN_2022}.
 
\begin{figure*}[tbh]
\includegraphics[width=340pt]{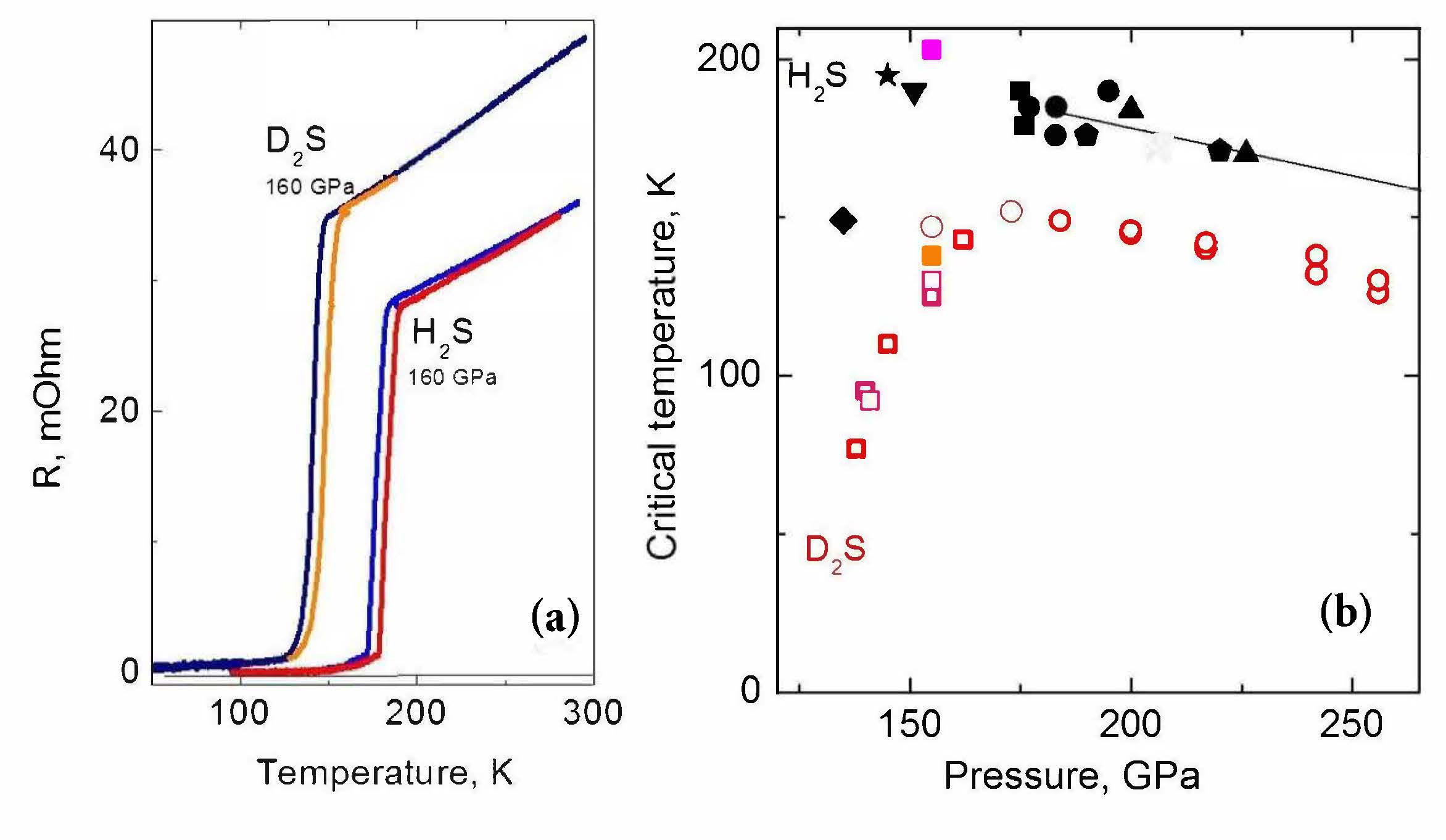}  
	\caption{Temperature dependences of resistance at the  superconducting transition (a)
		and 	pressure dependences of the critical temperature (b)  for
		 sulfur hydride and sulfur deuteride.
	Adapted from \cite{drozdov_Nat_2015}.
	}
	\label{fig:isotope}
\end{figure*}
As an example, figure \ref{fig:isotope} shows that the critical temperature decreases when hydrogen is replaced with a heavier element,
and this relationship holds in a wide range of pressures from 175 to 250\,GPa.

 \subsection{Influence of magnetic and nonmagnetic impurities on the superconducting critical temperature}
  \subsubsection{Magnetic impurities}
Introduction of impurities into a superconductor is an important tool for studying symmetry and the pairing mechanism. According to Anderson's theorem \cite{anderson_JPCS_1959, abrikosov_book},
non-magnetic impurities do not affect the isotropic singlet $s$-wave parameter
order in ordinary BCS-type superconductors \cite{bcs_PRB_1957},
whereas scattering from paramagnetic centers effectively destroys $s$-wave pairing 
\cite{abrikosov-gorkov_ZhETF_1960, abrikosov_book}.

In Ref.~\cite{semenok_AdvMat_2022} a series of ternary polyhydrides with the composition (La,\,Nd)H$_{10}$ was synthesized,
containing 8–20 at\% Nd. The  Nd$^{3+}$ ions have an outer electron shell of $4f^3$ and a magnetic moment of 3.62$\,\mu_B$/atom.
Since Nd atoms are randomly located in the lattice, they can be considered as paramagnetic impurities.
The main idea of this experiment was that Nd should effectively suppress superconductivity 
in LaH$_{10}$, while its structure {\it Fm\={3}m} remains practically unchanged due to the great similarity  between 
physical properties of La and Nd atoms.

For a small concentration of magnetic impurities, $x\ll 1$, the Abrikosov-Gorkov theory predicts a linear
dependence of $T_c$ on concentration $x$
   \cite{abrikosov-gorkov_ZhETF_1960, abrikosov_book}: 
\beq
T_c(0)-T_c(x)= \frac{\pi \hbar}{4 k_B\tau} x,
\label{eq:abrikosov-gorkov}
\eeq
where $\tau$ is the  collision time for scattering at a random impurity potential.
 
\begin{figure}[tbp]
	\includegraphics[width=130pt]{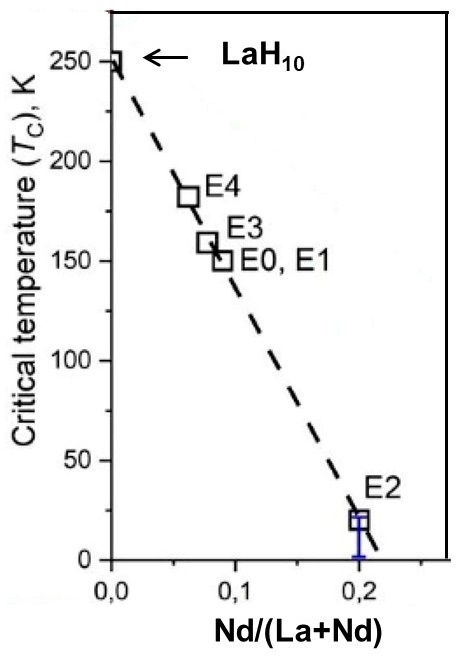}
	\caption{Dependence of the critical temperature for (La,\,Nd)H$_{10}$ on the relative concentration of Nd impurities.
The arrow marks the $T_c$ value for stoichiometric LaH$_{10}$.		Adapted from \cite{semenok_AdvMat_2022} 
	}
	\label{fig:Tc(x)} 
\end{figure}

In the case of (La,\,Nd)H$_{10}$, $\tau  \approx 5.4\times 10^{-15}$c \cite{semenok_AdvMat_2022},
and according Eq.~(\ref{eq:abrikosov-gorkov}),   each percentage of Nd impurity   should reduce 
$T_c$  in    LaH$_{10}$ by  $\Delta T_c\approx 10$\,K, or, in relative units,
$\Delta  T_c (1\% {\rm Nd})/T_c({\rm LaH}_{10}) = 0.044$.  
Comparing this result with experimental data, shown in Fig.~\ref{fig:Tc(x)}, we see  rather  good {\em  agreement} of the  theory predictions with experimental data.
It has been established that superconductivity completely disappears at $\approx 20$\% concentration of Nd impurities
(see Supplemental materials, Ref.~\cite{semenok_AdvMat_2022}).

To compare with conventional low-temperature BCS superconductors, we note
that the suppression of superconductivity in metallic La
with the introduction of magnetic  Eu and Gd impurities  was also studied in Ref. \cite{legvold_SSC_1976}. A quantitative agreement with the theory 
\cite{abrikosov-gorkov_ZhETF_1960} was found,
if one  takes into account corrections for the reduced scattering cross-section for Eu impurities compared to Gd impurities (due to the smaller exchange integral of the overlap of $4f-5d$ states).

\subsubsection{Nonmagnetic impurities}
 As for non-magnetic impurities, it is known that the introduction of a small concentration of carbon
 in La does not affect the critical temperature $T_c\approx 245$\,K of superconductivity in C:LaH$_{10}$
  (see a note in Ref.~\cite{semenok_AdvMat_2022}). Numerous experiments have  been performed with assembling the high-pressure diamond anvil cell in air, where an oxide film is inevitably present on
 surface of the loaded Y, La–Y, and La–Nd metals. Their results qualitatively confirm the absence of influence of non-magnetic 
 oxygen impurities  on the superconducting critical temperature of hydrides.

 Another example is the recent work on introducing sulfur into yttrium polyhydrides \cite{zhang_SciChPMA_2024}.
	As a result, the authors observed steps in the temperature dependence of electrical resistance corresponding to $T_c = 235$\,K and about $210-215$\,K. These values are consistent with previously obtained results for YH$_6$ and YH$_9$, indicating that non-magnetic sulfur does not influence the critical temperature of yttrium hydrides.
Indirect confirmation of the absence of influence of non-magnetic impurities are also almost identical 
values $T_c = (176 - 203)$\,K 
in the compounds $R\bar{3}m$-(La,\,Y)H$_{20}$ and $Pm\bar{3}m$-(La,\,Y)H$_{12}$ (see Table~2 in \cite{semenok_AdvMat_2022}),
which differ, in this context, in the concentration of nonmagnetic atoms.

 The most important effect of impurities on polyhydrides is a change in the region of their dynamic and thermodynamic stability. A striking example is the synthesis of lanthanum-cerium hydrides (La,\,Ce)H$_{9-10}$, which currently demonstrate the highest $T_c$ value at the lowest pressure: $T_c > 200$\,K at 100\, GPa \cite{bi_NatCom_2022, chen_NatCom_2023}.

Among pathological reports (about materials that are now 
called ``unidentified superconducting objects'', USO), we note recent announcement of the   ``miraculous'' impact of carbon doping on the superconductivity in H$_3$S \cite{snider_Nature_2020}.
Subsequent experimental works did not confirm this result \cite{goncharov_JAP_2022, sadakov_UFN_2022}, analysis of the experimental data themselves revealed their falsity \cite{vanderMarel-hirsch}
and the paper has been retracted  from the journal.

The  compound LuH$_x$N$_y$ is worthy of separate discussion. 
 The paper  reporting that this compound, when doped with non-magnetic nitrogen atoms, becomes superconducting with $T_c = 294$\,K at a pressure of only 10\,kBar \cite{NLuH} also turned out to be falsified. This report 
 was refuted by a number subsequent works and the corresponding article has been retracted \cite{retraction_note_NLu3H}. 
 
 However, what is interesting is the theoretically  predicted positive result of doping on the  $T_c$ value. 
 According to the band structure calculations  \cite{pavlov_JETPL_2023} it is achieved by introducing nitrogen atoms
 not into random, but into regular positions of the crystal lattice, replacing 1/4 of the hydrogen atoms. As a result of such special substitution, the Fermi level in LuH$_{2.75}$N$_{0.25}$ is lowered by $\approx 1.8$\,eV compared to LuH$_3$. Correspondingly, the density of states at the Fermi level almost doubles. It remains unclear, however, whether such a substitution is feasible in practice.

 \subsection{Meissner effect and diamagnetic screening}
Measurements of the electrical resistance vanishing  (Figs.~\ref{fig:R(T)_LaY}, \ref{fig:SnH4-R(T)}) are necessary but not sufficient to prove the existence of superconductivity.
 In addition to them, it is necessary to demonstrate directly  the effect of the external magnetic field expulsion from the sample volume (Meissner effect). However, measuring  Meissner effect using a SQUID magnetometer or the inductive tools at pressures above 13\,GPa is difficult, since usually the signal from a sample located in a diamond anvil cell is  orders of magnitude smaller than the signal from the materials from which the cell and the gasket are made.
Nevertheless, several experiments  aimed on observing diamagnetic screening in hydrides have been performed using  SQUID magnetometry 
  \cite{eremets_UFN_2016, huang_NatSciRev_2019, struzhkin_MatRadEx_2020, minkov_NatCom_2022, semenok_NatSciRev_2019, struzhkin_Science_2016}, AC magnetic susceptibility measurements  \cite{huang_NatSciRev_2019, struzhkin_MatRadEx_2020}, and M\"{o}ssbauer spectroscopy  \cite{troyan_Science_2016}.  

In all experiments so far, it has been possible to observe only the absence of a magnetic field inside the sample 
cooled in zero field (ZFC mode). Obviously, this is a manifestation of the ``diamagnetic sceening'' effect.
For the reason noted above, it has not yet been possible to observe the true expulsion of a magnetic field from a sample when it is cooled in the presence of a field (FC) \cite{Hirsch_JPhysC_2021}.
It is worthnoting,   while the superconducting transition is clearly visible in the ZFC measurements, its signature is barely or almost  not visible in measurements with cooling in a magnetic field (FC) \cite{minkov_NatCom_2022}. 

In addition to the technical challenges associated with magnetic measurements in diamond anvil cell,
there are also features in the expulsion of weak magnetic flux in type II superconductors, associated
with strong vortex pinning \cite{tomioka_PhysC_1993}. Strong pinning prevents vortices from moving inward
and outward  the  sample interior in fields below  $H_{c1}(T)$. 
As shown in Refs. \cite{gokhfeld_JAP_2011, gokhfeld_JTPL_2019},
in a type II superconductor (which SC hydrides certainly are), Abrikosov vortices and a magnetic field in the center of the sample are absent as long as the external field is less than the full penetration field $H_p$. A corresponding analysis of this effect for specific measurements with H$_3$S was carried out in \cite{troyan_UFN_2022}.  It was shown that in the field range $H_{c1} < H < H_p$, the distribution of Abrikosov vortices in the sample is inhomogeneous, like in all type II superconductors, where the magnetic flux density decreases from the  sample edges to the center. Therefore, the formulas for a uniform field are not applicable to such experiments, because use an overestimated $H_{c1}$ value. The values of $H_{c1}$ and critical current $j_c$ for H$_3$S obtained as a result of the analysis \cite{troyan_UFN_2022} are quite consistent with similar parameters for other type II superconductors. 

In Ref.~\cite{troyan_Science_2016}
a different technique was used to detect the diamagnetic properties of superconducting hydrogen sulfide at high pressures, namely M\"{o}ssbauer spectroscopy \cite{lyubutin_book, lyubutin_JETPL_1975}. The magnetic field detector was a thin tin foil enriched with the M\"{o}ssbauer Sn-119 isotope. The magnetic moments of Sn-119 nuclei are an order of magnitude greater than those of the traditionally used M\"{o}ssbauer iron isotope Fe-57, so Sn-119 nuclei are more sensitive to the magnetic environment than Fe-57 nuclei. In nuclear $\gamma-$resonance, M\"{o}ssbauer spectra were registered during transitions between the nuclear levels of the ground and excited states of Sn-119  with spin 1/2 and 3/2, respectively. 

Synchrotron experiments were performed in the nuclear resonance forward scattering (NRS or NFS) mode for two magnetic field directions — parallel and perpendicular to the sample plane. In this mode, synchrotron radiation consists of picosecond pulses, the time interval between which is 800\,ns or more.
During this period, the time decay  of radiation from the nuclei of the M\"{o}ssbauer isotope after pulsed resonant excitation is recorded. The shape of the spectra depends on the magnetic state of the sample. In the absence of a magnetic field, the nuclear scattering signal decays exponentially. In the presence of a magnetic field, so-called quantum beats appear, caused by the interference of radiation during transitions between the ground and excited states of Sn-119 nuclei split by the magnetic field. In the spectra of resonant nuclear scattering, this manifests 
in the form of oscillations of the signal amplitude.

In these measurements  \cite{troyan_Science_2016}, the M\"{o}ssbauer sensor showed the magnitude of the magnetic field that penetrated the sample at a given temperature. It was found that in the temperature range 4.7 - 90\,K the superconductor H$_3$S completely screens the M\"{o}ssbauer sensor from magnetic field. Above this temperature, external magnetic field partially penetrates the sample, however, full penetration of the field occurs only above 145\,K. The data obtained confirm the effect of diamagnetic screening in H$_3$S of a magnetic field of 0.7\,T up to temperatures of 90-100\,K. Partial screening of the magnetic field is maintained up to approximately 145\,K. This confirms that sulfur hydride H$_3$S, compressed to 150\,GPa, is a type II superconductor with very high critical parameters.

When analyzing the results of studying diamagnetic screening in LaH$_{10}$ and H$_3$S samples using a SQUID magnetometer, it is necessary to take into account that the hydride samples are probably porous and consist of microscopic grains of  $\sim 0.05-0.5\,\mu$m size. In such case, the demagnetization factor $N$ calculated for a random packing of spherical particles 
ranges from 0.33 to 0.5 \cite{bjork_APL_2013, prozorov_PRA_2018}.
External magnetic field penetrates the sample between individual grains and, therefore, no change in the magnetization of the sample is observed at temperatures around $T_c$ upon cooling in the field (FC). Thus, the found values of the penetration field $H_p(0) = 96$\,mT for H$_3$S and 41\,mT for LaH$_{10}$ are the lower bound of $H_{c1}(0)$, whereas a more realistic estimate gives  $H_{c1}(0) \sim H_p(0)/(1-N) = (1.5-2)\times H_p(0)$.

\section{Unusual transport properties of hydrides in the S and N states}

\subsection{Upper critical field}
In the Ginzburg-Landau theory \cite{GL_ZhETF_1950} the upper critical field 
 \beq
 H_{c2}(T=0) =\frac{\phi_0}{2\pi \xi_0^2},
 \eeq
where $\phi_0$ is the magnetic flux quantum and $\xi_0$ the coherence length.
  
\begin{figure}[tbp]
\includegraphics[width=220pt]{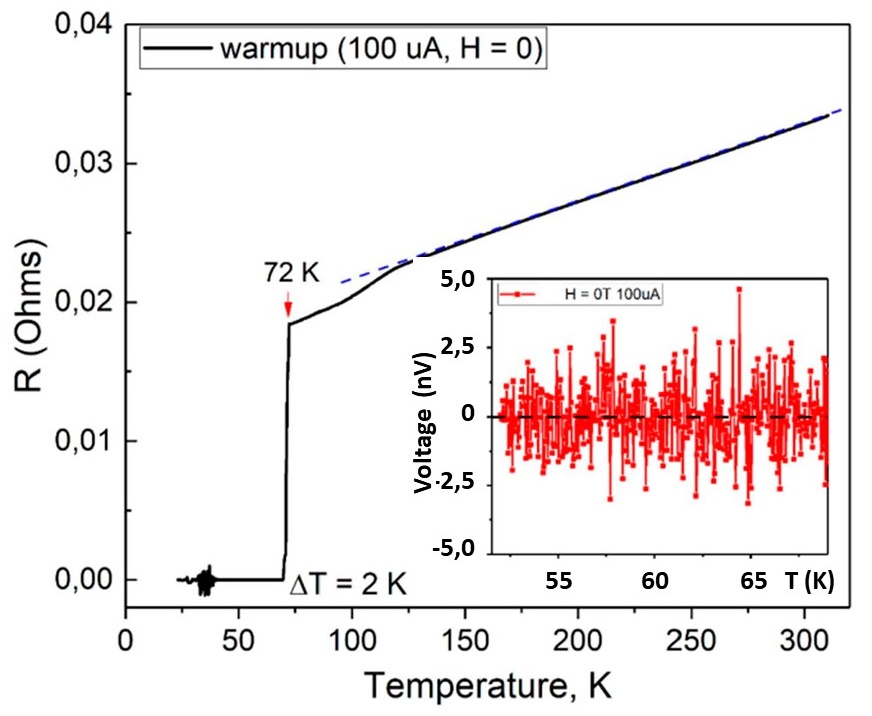} 
	\caption{Superconducting transition $R(T)$ curve  for tin hydride at 180\,GPa.  
The insert shows  voltage drop between the  potential contacts to the sample, 
 measured  in its superconducting state with a bias current of $100\,\mu$A. Adapted from Ref.~\cite{troyan_AdvSci_2023}.
}
	\label{fig:SnH4-R(T)}
\end{figure}

\begin{figure}[tbp]
	\includegraphics[width=240pt]{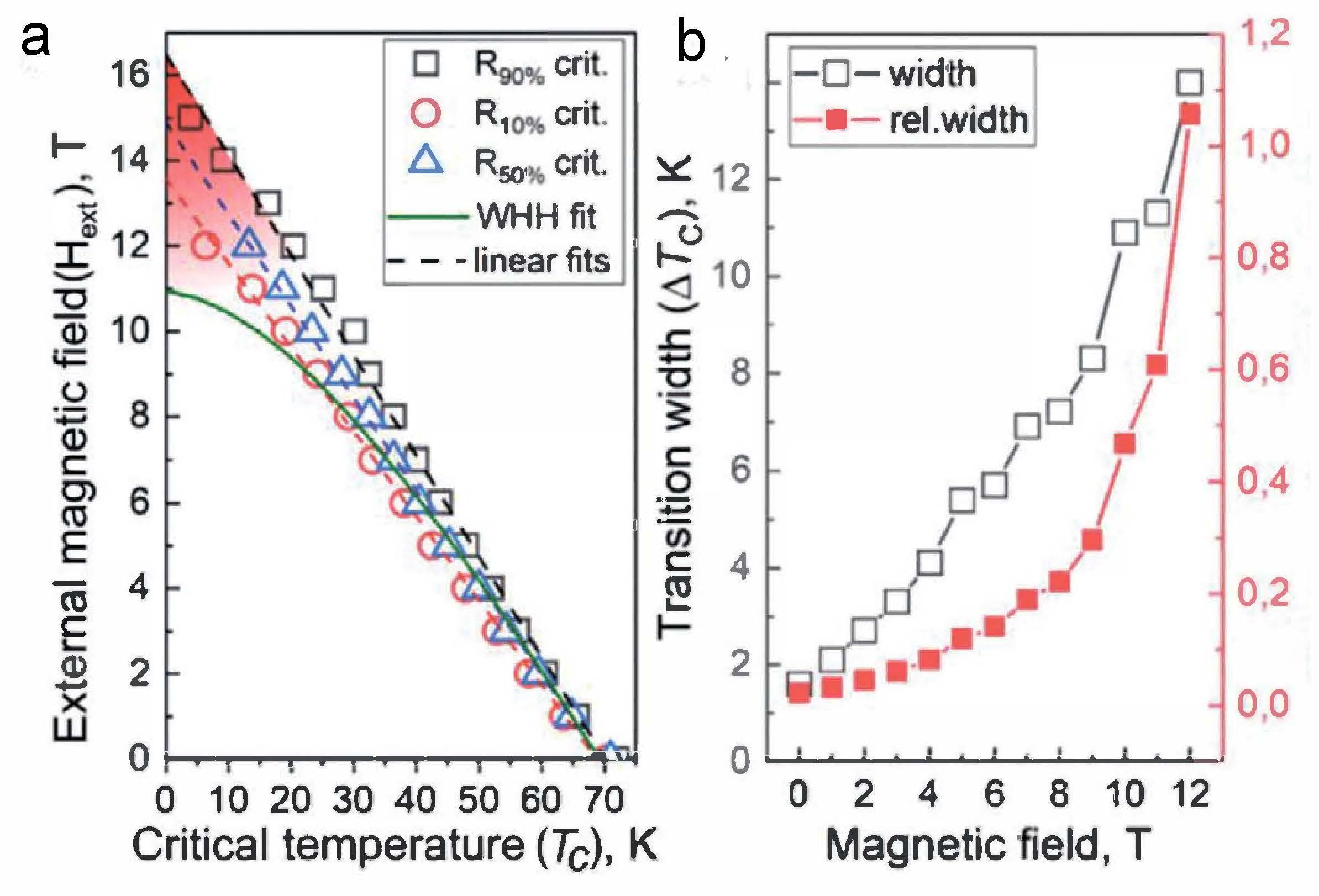}  
	\caption{(a) Dependence of the upper critical field on temperature for SnH$_4$, plotted using three different criteria for determining the  $H_{c2}$ value. The solid line shows the approximation using the WHH formula  \cite{WHH}.  (b) Dependence of the transition width $\Delta T_c$ and the relative broadening of the transition $\Delta T_c/T_c$ on the magnetic field. Adapted from Ref.~\cite{troyan_AdvSci_2023}.
	}
	\label{fig:Hc2(T)}
\end{figure}

Due to the extremely high values of the upper critical field $H_{c2}(T=0)$, the influence of  magnetic field on the superconducting transition in hydrides can be traced, as a rule, only in the region of high temperatures, near $T_c$. In order to study the  $H_{c2}(T)$ dependence  in a wider range of the normalized values of $T/T_c$ in Ref.~\cite{troyan_AdvSci_2023}, 
 SnH$_4$ compound with a relatively low  $T_c\approx 72$\,K value as been chosen (see Figs.~\ref{fig:SnH4-R(T)}) and \ref{fig:Hc2(T)}). A possible reason for such a low $T_c$ value in  SnH$_4$ is 
a low density of electronic  states at the Fermi level \cite{troyan_AdvSci_2023}.
 
The temperature dependence of $H_{c2}$, measured in steady fields of a superconducting magnet, is shown in Fig.~\ref{fig:Hc2(T)}a. At $T\rightarrow 0$ this dependence is extrapolated
to $H_{c2}(T=0) \approx 16$\,T; such a low value of $H_{c2}(0)$ made it possible to measure this dependence over the entire range of fields, from 0 to $H_{c2}(T=0)$.
The broadening of the transition with temperature, shown in Fig.~\ref{fig:Hc2(T)}b, illustrates what was said above: the broadening  $\Delta T_c/T_c$ in low fields changes weakly, but then increases sharply  as the field increases.
  
The most interesting and unusual result is the functional $H_{c2}(T)$ dependence: it is almost linear over the entire temperature range up to $T_c$. When measuring in a pulsed field up to 68\.T, a linear
dependence $H_{c2}(T) \propto (T_c - T)$  was also observed in Ref. \cite{semenok_AdvMat_2022} for (La,Nd)H$_{10}$. For superconductors described by the Bardeen-Cooper-Schrieffer (BCS) theory, the generally accepted model for the $H_{c2}(T)$  dependence is
the Werthamer–Gelfand–Hohenberg (WHH) model, which predicts a flattening of the $H_{c2}(T)$ dependence at low temperatures  \cite{WHH}.

\begin{figure}[tbp]
		\includegraphics[width=210pt]{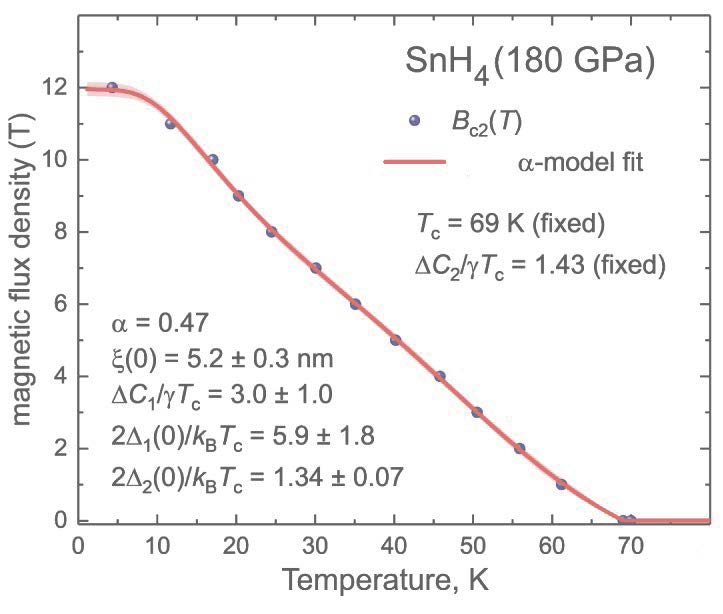}  
		\caption{An example of approximation of the measured  $H_{c2}(T)$ dependence within the framework of the  two-component alpha models of SP condensate. The model parameters are shown in the figure. Adapted from 
			Ref.~\cite{troyan_AdvSci_2023}, Supplemental information.
		}
		\label{fit_Hc2(T)}
\end{figure}

The linear dependence $H_{c2}(T)$ is inherent not only to SnH$_4$ and (La,Nd)H$_{10}$, it is also observed in many other polyhydrides, for example, YH$_4$, LaH$_{x}$, etc. A similar linear or quasi-linear dependence of $H_{c2}(T)$ was also observed in iron pnictides \cite{hunte_Nat_2008, yuan_Nat_2006, khim_PRB_2011}. In a number of cases it could be explained by the presence of several superconducting gaps in the spectrum  \cite{ummarino-bianconi_CondMat_2023, hunte_Nat_2008, yuan_Nat_2006, khim_PRB_2011}. An example of an attempt to describe the measured dependence $H_{c2}(T)$ within the framework of a two-component, the so-called ``alpha model'' of a superconducting condensate is shown in Fig.\ref{fit_Hc2(T)}  (see \cite{troyan_AdvSci_2023}, Supplementary Information).

Indeed, for most polyhydrides, due to the large number of electrons per unit cell, many bands are filled and the Fermi surface is multiband. Besides, for a number of polyhydrides (LaH$_{10}$, YH$_{10}$, YH$_9$) the existence of a two-component SC condensate was predicted theoretically  \cite{wang_PRB_2020, kuroki_PRB_2005, semenok_thesis}. However, this explanation is not universal, since superconductivity, for example, in CaH$_6$ is single-gap, according to the theoretical results 
\cite{jeon_NJP_2022}. 

An alternative explanation for the linear  $H_{c2}(T)$ dependence may be the mesoscopic inhomogeneity of the samples, in the volume of which there are regions with slightly different compositions
and different values of $T_c$ and $H_{c2}$ \cite{semenok_AdvMat_2022, spivak_PRL_1995, galitski_PRL_2001}.
Indeed, despite the fact that the sharp drop in resistance at the superconducting transition, as well as the  X-ray analysis indicate the macroscopic homogeneity of the SP hydrides, the existence of inhomogeneities on
mesoscopic scales cannot be excluded. 
Theoretical models do not fully explain the linear dependence of $H_{c2}(T)$. Moreover, in the work \cite{spivak_PRL_1995}
some straightening of the standard BCS dependence is predicted due to the appearance of a section with positive curvature on $T_c(H)$.
The model \cite{galitski_PRL_2001} predicts an increase in $H_{c2}$ at $T\rightarrow 0$. As long as the superconducting regions in the sample volume are interconnected with Josephson tunnel coupling, superconductivity will manifest itself in the sample volume.

The linear dependence $H_{c2}(T)$ was previously observed in InO films, and to explain it in the work \cite{sacepe_NatPhys_2019} it was proposed that the SC state is a vortex glass state, thermal fluctuations in which lead to such a dependence. For SC hydrides, scaling analysis of the critical current in ThH$_{10}$ \cite{sadakov_2311.01318} revealed the dependence $j_c\propto (1-T/T_g)^{1.6}$, where $T_g$ was interpreted as the transition temperature into a vortex glass state. The dependence with such an exponent does not contradict the result of the Ginzburg-Landau theory $j_c\propto \rho_s/\xi_{\rm GL} \propto (1-T/T_c)^{3/2}$, however, such an interpretation and its applicability to the results for polyhydrides require more detailed study. 

We note also that for a typical hydride CeH$_{9-10}$ the Fermi temperature is $T_F \sim 6.5\times 10^4$\,K.
Therefore, the ratio $T_c/T_F \sim 1.5\times 10^{-3}$ is not small, unlike
simple superconducting metals (e.g.,  Sn, In, etc.), for which this ratio is $\sim 10^{-5}$.
For hydrides with higher $T_c$ values, the ratio $T_c/T_F$ is rather close
to superconductors based on iron pnictides and cuprates.
Likewise, the ratio $2 \Delta(0)/T_c \approx 4$ \cite{talantsev} is also not small.
For these reasons, SC hydrides should be considered as  superconductors with moderately strong coupling. 

\subsection{Linear temperature dependence of the normal state resistance }
In the normal state, the transport properties of SP hydrides are also not yet fully understood.
For many hydrides, in a wide temperature range $T > T_c$ in the absence of a magnetic field, a linear temperature dependence of resistance is observed: this dependence is visible in Fig.~\ref{fig:SnH4-R(T)} in the range $T=120 - 320$ \,K.

A similar temperature dependence of $R(T)$ was observed for LaH$_{10}$ \cite{semenok_AdvMat_2022, sun_NatCom_2021}, for CeH$_{9 - 10}$ \cite{semenok_2307.11742} and for a number of other hydrides. 
In all cases it is linear, for example for CeH$_{9 - 10}$ - in the range $\sim 110-300$K \cite{semenok_2307.11742}. The linear dependence of the resistance cannot be approximated with the Bloch-Gr\"{u}neisen dependence \cite{bloch_1930} for electron-phonon scattering. Indeed, an attempt to use such fitting for SnH$_4$ leads to an unrealistically low Debye  temperature value $\theta_D \approx 100$\,K \cite{troyan_AdvSci_2023}, which contradicts the phonon spectra of hydrides with powerful vibrational peaks of hydrogen atoms at high frequencies. 

Talantsev \cite{talantsev} successfully approximated $\rho(T)$ for (Ln,\,Nd)H$_{10}$ with the $T^5$ dependence and obtained a plausible estimate  $\theta_D =1150$\,K. However, the experimental data $\rho(T)$ for this compound differed only slightly from the linear dependence, which makes the approximation result unreliable. We note that in many other cases (an example is shown in Fig.~\ref{fig:SnH4-R(T)}) the difference between the measured $R(T)$ and the linear function is even smaller. For example, for SnH$_4$, when approximating the experimental data $\rho(T)$ with the function $R=R_0+AT^n$,  the fitting parameter $n$ in Ref.~\cite{troyan_AdvSci_2023} was determined much lower ($n=0.9$), than the expected value  $n=5$ (see Fig.~22, Supplemental materials, \cite{troyan_AdvSci_2023}). 

 Moreover, if the 	$R(T)$  dependence
 in a zero magnetic field arose due to scattering by phonons and was described by the Bloch-Gr?neisen
  formula, then the application of a magnetic field would not have any effect on it. 
	Indeed, magnetic field does not change either the phonon spectrum or the matrix element 
	of electron-phonon scattering.
	However, it was found experimentally \cite{semenok_tbp}, that application of  20\,T field to 
(La,\,Ce)H$_{10}$ at 148\,GPa, 
changes the situation and ``straightens'' the $R(T)$ dependence (makes it linear), eliminating the chances of the $\propto T^5$ approximation.
 This result (straightenning of $R(T, H\neq 0)$), therefore,  indicates a non-phonon mechanism of the linear  $R(T)$ dependence. 

In general, not many physical mechanisms are known that lead to a linear temperature dependence of the resistance of metals.
To assess their applicability, we note that the hydrides under consideration in the normal state have a Fermi energy $T_F \sim (3 - 10)\times 10^4$\,K, carrier density $n=(20-60)\times 10^{21}$/cm$^3$,
and are metals, $ E_F\tau/\hbar \gg 1$. The dimensionless electron-electron interaction parameter $r_s=E_{\rm ee}/E_F$ in the normal state for hydrides is not small, for example for CeH$_9$ \cite{semenok_2307.11742}
it is $r_s \approx 2.5$.
For a strongly correlated normal metal, a positive temperature dependence $d\rho/dT>0$, in principle, can arise due to scattering by impurities taking into account the electron-electron interaction.
For the concentration estimated above, the Fermi-liquid coupling constant is $F^a_0 \approx -0.2$,
therefore, Fermi-liquid effects should not be small \cite{CDFLM}.
 
However, only for the two-dimensional case they lead to
linear dependence $\rho(T) \propto T$ \cite{ZNA} and only in the ballistic  interaction regime, $k_BT\tau/\hbar\gg 1$. For the three-dimensional case the interaction effects lead to the dependence $\propto T^{1 /2}$ \cite{CDFLM}, which does not correspond to the observed linear one. 

For generality, we note that the linear dependence $\rho(T)$ exists in the normal state not only in hydrides, but also in other HTSC materials - iron pnictides (FeSe$_{1-y}$S$_y$), La nickelates La$_3$Ni$_2$O$_7$ \cite{zhang_2307.14819} and cuprates (La$_{2-x}$Sr$_x$CuO$_4$) \cite{cooper_Science_2009}. In all cases,  this dependence  has not yet found a satisfactory explanation.

\subsection{Linear magnetoresistance}
\begin{figure*}[tbp]
	\includegraphics[width=170pt]{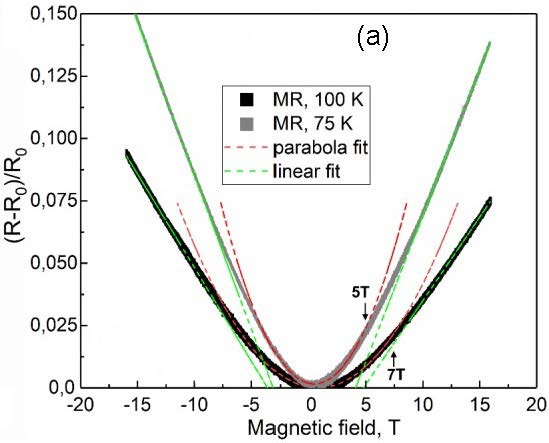}  
	\includegraphics[width=152pt]{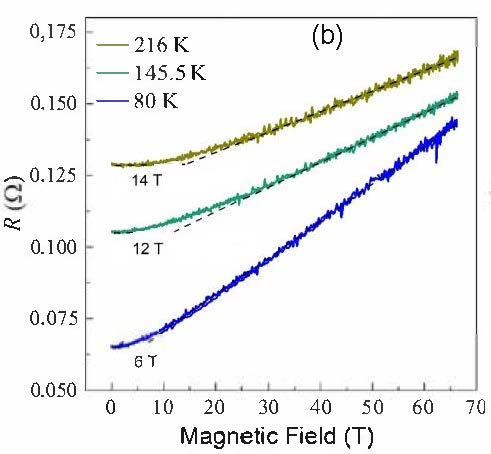}  
	\includegraphics[width=152pt]{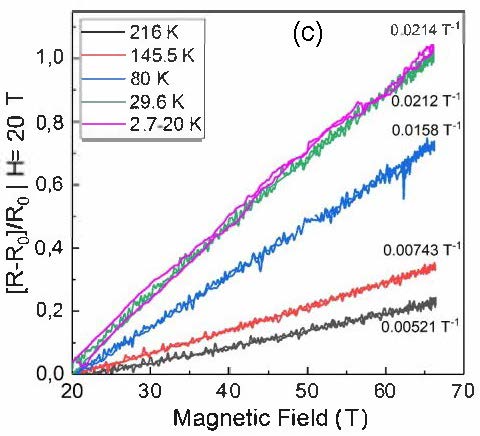}  
	\caption{Dependence of the normal state  resistance for SnH$_4$ on the magnetic field at 180\,GPa: (a) $R(H)$ in weak fields up to 16\,T at $T= 75$ and 100\,K; (b) $R(H)$ in the normal state at $T > 70$\,K; (c) the linear part of the resistance normalized to the  zero field value, plotted in the fields above 20\,T.The numbers next to the curves indicate the slope $\Delta R(H)/(R_0 H)$.
		Due to overheating by eddy currents in a pulsed field, the lowest temperature of the sample reaches $\approx 20$\,K. Adapted from Ref.~\cite{troyan_AdvSci_2023}.
	}
	\label{fig:SnH4-R(H)}
\end{figure*}

In the normal state of  many  polyhydrides, in weak magnetic fields, the electrical resistance increases quadratically with increasing field (see Fig.~\ref{fig:SnH4-R(H)}),
which is typical for a multiband metal (one could even say that such magnetoresistance indicates a multiband Fermi surface). However, as field  increase further, this dependence changes and the resistance starts growing linearly and continues to grow so, up to the maximum fields attainable in the laboratory. 

Linear dependence of magnetoresistance on magnetic ield 
was discovered by P.~L.~Kapitsa in 1929 \cite{kapitza} for polycrystalline samples. Dreizin and Dykhne \cite{dreizin} explained the linear magnetoresistance by taking into account scattering at crystallite boundaries in strong magnetic fields $\omega_c\tau \gg 1$.
In addition, Kapitza's law in polycrystals arises as a result of angle averaging  of the polar magnetoresistance diagram for the metals that have open Fermi surface (FS) sections (for example, Cu, Ag, Au, In, Pb).
The presence of open sections of the FS of hydrides, indeed, follows from  the band structure calculations of several  hydrides, e,g.,
  LaH$_{10}$, and YH$_6$ \cite{heil_PRB_2019, liu_PRB_2019}, but the open sections of FS have not yet been identified experimentally.

Apart from hydrides, 
the linear dependence of $R(H)$ is also observed in quasi-two-dimensional ``bad metals'' - SrZnSb$_2$ \cite{wang_APL_2012}, in semimetals - Ni$_3$In$_2$S$_2$ \cite{fang_preprint}, as well as in ferromagnetic MnBi \cite{he_NatCom_2021}.
 Compounds with charge density wave (CDW) state  and with incomplete Fermi surface nesting also exhibit linear magnetoresistance at temperatures below the CDW critical temperature: quasi-1D compounds   (e.g.,  NbSe$_3$ \cite{richard_PRB_1987}),  quasi-2D transition metals dichalcogenides  (such as  2H-NbSe$_2$ and 2H-TaSe$_2$ \cite{naitol_JPSJ_1982})  and  also rare-earth tritellurides (e.g., TbTe$_3$ and HoTe$_3$ \cite{sinchenko_PRB_2017}).
This dependence is associated with the scattering of charge carriers by fluctuations of the CDW order parameter \cite{sinchenko_PRB_2017}.

 Polyhydrides in the normal state, however, are the good metals, and have neither a two-dimensional spectrum nor ferromagnetism.
 The temperature dependence of the resistance of polyhydrides does not show signs of a transition to the CDW state.
Finally, linear magnetoresistance is observed in materials with a Dirac spectrum, for example, in graphene, in gapless semiconductors or semimetals with  a very small carrier
concentration \cite{abrik_PRB_1998, abrik_PRB_1999}. However, such a spectrum and small carrier density are also not inherent in polyhydrides. 

Recently, it was found that many polyhydrides, such as La$_4$H$_{23}$
	\cite{guo_NatSciRev_2024}, 	CeH$_{10}$ \cite{arxiv2307.11742}, ThH$_9$ and (La,\,Ce)H$_{10}$
	exhibit negative magnetoresistance above the superconducting transition in strong magnetic fields. It can be assumed that this behavior is related with the presence of a pseudogap state in hydrides, just as is observed in cuprate superconductors.

\section{Conclusion}
Superconducting polyhydrides with critical temperatures around the room temperature values, as very ``young'' materials,
attract a great deal of interest of researchers.
The most obvious and easily interpretable are the results of transport and magnetic properties measurements.
Numerous galvanomagnetic measurements have documented and reproduced the  drop in resistance
at a temperature below the critical value $T_c$, indicating a superconducting transition. With increasing external magnetic field the critical temperature decreases and the superconducting transition broadens, which is also consistent with this interpretation.

In magnetic measurements, static diamagnetic screening has also been  observed in numerous experiments, where an external field was applied to a sample cooled in zero field (ZFC). These experiments were carried out using measurements of the magnetic moment and magnetic susceptibility with a SQUID magnetometer and with the M\"{o}ssbauer spectroscopy using synchrotron $\gamma$ quanta radiation.

Due to the technical difficulties of measuring a small signal, unfortunately, it has not yet been possible to reliably detect the expulsion of a magnetic field from a sample volume  when it is cooled in a magnetic field (FC), i.e. the true Meissner effect. Such measurements would be very important in proving a truly superconducting state.
The observation of the isotope effect in superconducting hydrides is convincing evidence of the electron-phonon mechanism of electron pairing. Experiments on the influence of scattering by magnetic impurities on the critical temperature
are in agreement with the Abrikosov-Larkin theory and prove the singlet nature of superconducting pairing.

As a consequence of the above experimental results, independently reproduced in several laboratories, it is now generally accepted that hydrides are ordinary superconductors with singlet pairing and moderately strong coupling. Until recently it was believed  that their behavior in the normal state can be approximately described within the framework of the ordinary Fermi liquid model.
However,  accumulating experimental data casts doubt on this point of view. The most obvious contradictions with the properties of BCS superconductors and normal metals are those presented in this article: (i) linear temperature dependence of the second critical field, (ii) linear temperature dependence of resistance in the normal state,  (iii) linear positive magnetoresistance in a strong magnetic field $\omega_c\tau\gg 1$,  and (iv) negative magnetoresistance in strong magnetic fields.

Each of the above listed effects, in principle, was encountered previously for different classes of materials and found its own individual explanation. However, in the aggregate, these anomalous properties are found only in cuprate HTSCs \cite{legros_NatPhys_2019, ataei_NatPhys_2022, greene_ARCMP_2020} and have not yet found a satisfactory microscopic explanation. The  state at $T>Tc$ in cuprates is phenomenologically associated with the so-called ``strange metal'', and the superconducting state at $T<T_c$ - with a superconductor with moderately strong coupling.

Progress in the synthesis of new superconducting hydrides is happening so quickly that it is worth asking whether there is a limit for increasing the critical temperature of superconductivity. At the beginning of the 21st century V.~L.~Ginzburg  answered this question negatively, meaning the absence of theoretical restrictions on the critical temperature reaching
293\,K.
The data accumulated to date allows for a more detailed assessment. Within the framework of the Eliashberg-McMillan theory for dirty superconductors with strong coupling and a phonon pairing mechanism
the critical temperature depends on 3 parameters - the ``average'' phonon frequency $\langle \omega_{\rm log}\rangle$, the electron-phonon interaction constant $\lambda$ and the Coulomb pseudopotential $\mu^*$.
According to McMillan's semi-empirical formula
\beq
k_B T_c \approx \frac{\hbar\omega_{\rm log}}{1.2}\exp\left[-\frac{1.04(1+\lambda)}{\lambda-\mu^*(1+0.62\lambda)}\right].
\label{eq:McMillan}
\eeq

With two correction functions $f_1, f_2(\lambda, \omega_{\rm log},\omega_2,\mu)$ this Allen and Dines refined formula for the regime of moderately strong coupling $\lambda <1.5$ looks as follows \cite{allen-dynes}:

\beq
k_B T_c \approx f_1 f_2 \frac{\hbar\omega_{\rm log}}{1.2}\exp\left[-\frac{1.04(1+\lambda)}{\lambda-\mu^*(1+0.62\lambda) }   \right],
\label{eq:Tc}
\eeq
 where $\omega_{\rm log}$ is the logarithmic average frequency, and $\omega_2$ - the mean square frequency.
The Coulomb pseudopotential $\mu^*$ in the case of strong coupling decreases approximately by half 
due to the weakening of the Coulomb interaction, described by  the so-called Tolmachev logarithm
\cite{tyablikov_ZhETF_1958, kresin_book}
\beq
\mu^*= \frac{\mu}{1+\mu \ln (E_F/\hbar\omega_D)},
\eeq
where $\mu$ is the averaged potential of Coulomb interaction  of electrons in a metal
and $\omega_D$ – the characteristic phonon energy (for example, Debye frequency). As a result, 
$\mu^*$ takes values of $\approx 0.1 - 0.15$ (which are determined by numerical calculations).

The electron-phonon interaction constant $\lambda$,  the average logarithmic frequency, and  the mean square frequency
\bea
\lambda &=& 2\int_0^\infty \frac{\alpha^2 F(\omega)}{\omega}d\omega \nonumber\\
\omega_{\rm log} &=& \exp\left[\frac{2}{\lambda}\int_0^\infty \frac{\alpha^2 F(\omega)}{\omega}\ln\omega\,d\omega\right] \nonumber\\
\omega_2 &=&\sqrt{\frac{1}{\lambda}\int_0^{\omega_{\rm max}}\left[ \frac{2\alpha^2 F(\omega)}{\omega}  \right]\omega^2d\omega }
\label{eq:Tc-omega}
\eea
are calculated through the Eliashberg spectral function 
 $\alpha^2 F(\omega)$. 

As can be seen from (\ref{eq:Tc-omega}), the most important parameter is the electron-phonon interaction constant $\lambda$.
Hypothetically, in the regime of extremely strong coupling $\lambda \gg 1$, the exponential dependence $T_c (\lambda)$ (\ref{eq:Tc}) should turn into the root dependence $T_c \sim \lambda^{1/2}\tilde{\omega}$ \cite{kresin_book}, where $\tilde{\omega} =\langle \omega^2 \rangle^{1/2}$. However, the maximum value of $\lambda$ may be limited by the stability of the system and the translational invariance of the lattice.

Possible restrictions on the maximum value of $\lambda_{\rm max}$ associated with
violation of the adiabatic approximation, Migdal's theorem and the Migdal-Eliashberg theory as a whole \cite{sadovskii_JSNM_2020} have been discussed many times in the literature.
Initially, within the framework of the Fr\"{o}hlich Hamiltonian, the constraint $\lambda <0.5$ was obtained to ensure the stability of the phonon spectrum (positive phonon frequency). However, this limitation was obtained for the nonadiabatic case
$\hbar\tilde{\omega} \gg E_F$ \cite{sadovskii_JSNM_2020}, which is irrelevant to most superconductors.
Similarly, bipolaron instability at $\lambda \simeq 1$ \cite{alexandrov_UFN_1992} corresponds to the nonadiabatic case.
 
 Another constraint $\lambda =2$, at first glance, naturally arises from maximizing $T_c$ according to McMillan's formula (\ref{eq:McMillan}), since the maximum of $T_c$ [i.e. $\partial T_c/\partial \lambda =0 $] is achieved precisely at $\lambda =2$. However, this limitation is also apparent, since the formula itself is valid only for $\lambda \leq 1.5$  \cite{kresin_book}. To date, for many polyhydrides, numerical calculations have determined
 significantly larger values of $\lambda$, for example
 1.84 - 2.3 (for H$_3$S, depending on the pressure value) \cite{errea_Nature_2016, talantsev_SUST_2021},
 2.06 (for LaH$_{10}$) \cite{peng_PRL_2017},     2.41 (for YH$_{10}$) \cite{peng_PRL_2017}, 
2.76 (for LaH$_{10}$) \cite{errea_Nature_2020, talantsev_SUST_2020},  and,  finally, 
3.87 (also for (La,\,Y)H$_{10}$) \cite{song_ChemMat_2021}.

Within the same framework, it would be possible to tackle the inverse problem - 
optimization of the  Eliashberg spectral function. The fact is that in hydrides
$\alpha^2F(\omega)$ has two powerful peaks:
the peak at low frequencies, $\omega_1 \sim (5 - 10)$\,THz,
is associated with acoustic vibrations of metal atoms and has almost no effect on the value of $T_c$, and the high-frequency peak, $\omega_2 \sim 60$\,THz, is associated with hydrogen vibration modes \cite{pickard_ARCMP_2020, pickett_RMP_2023}.
In this context, high pressure helps to increase the vibrational frequency of hydrogen atoms. The calculated values of the average logarithmic frequency for known polyhydrides are 1080\,K (H$_3$S), 1340\,K (YH$_{10}$), 1210\, (ThH$_{10}$), and
1330\,K (YH$_6$) \cite{semenok_thesis}. The extended intermediate spectral region  is often empty, which negatively affects the values of $\omega_{\rm log}$ and $T_c$. 

 As an example of the reverse engineering of an effective spectral function, consider the model rectangular function $\alpha^2 F(\omega) = Const =a$ in the frequency interval from $\omega_{(1)}$
up to $\omega_{(2)}$ and equal to 0 outside this interval. Then $\omega_{\rm log} = (\omega_{(1)} \omega_{(2)})^{1/2}$
	and  $\lambda = 2a \ln\left( \omega_{(2)}/\omega_{(1)}\right) =3.6a \leq 3.6$.

According to modern theory, the value of $\lambda_{\rm max}$ is limited by the violation of translational symmetry of the lattice and the formation of a gap near the Fermi level \cite{esterlis-kivelson_PRB_2018, altshuler_PRB_2022}.
In the latter work, the most ``optimistic'' numerical estimate of the value of $\lambda_{\rm max} \approx 3.0 - 3.7$ was obtained, above which the lattice loses stability. As can be seen, this estimate translates the problem under consideration
far beyond the standard BCS theory of weak coupling $\lambda \ll 1$. 

From the brief historical consideration given above, it is clear that in experimentally discovered
new superconductors, the coupling constant $\lambda$ repeatedly exceeded theoretical limits,
which turned out to be associated with the limited applicability of the models.
Note that even the value $\lambda \approx 3$ is sufficient to obtain superconductivity at room temperature.

Regarding the maximum possible phonon frequency, one can also give an estimate based on the maximum speed of sound in
crystals \cite{brazhkin}
\beq
\frac{v_s}{c} = \alpha \left(\frac{m_e}{2m_p} \right)^{1/2}
\eeq
where $\alpha$ is the fine structure constant, $m_e,\, m_p$ -  the electron and proton masses, respectively.
From here we get $v_s <36.1\times 10^5$cm/s \cite{brazhkin}, and an estimate for the maximum value of $\omega_{\rm log} \sim 2500$K.
Using these estimates for the maximum possible parameters, we obtain a rough estimate for $T_c^{\rm max} \gtrsim 600$\,K.

\section{Acknowledgments}
I.~T. was partially supported by the RSCF grant 
22-12-00163. D.~S. thanks National Natural Science Foundation of China (NSFC, grant No. 1231101238) and Beijing Natural Science Foundation (grant No. IS23017) for support of this research.
 Measurements were taken using equipment of  the Shared Facility Center
 of the Lebedev Physical Institute. The research was carried out within the framework of state assignments from the Scientific Research Center ``Kuchatov Institute''  and LPI.

\end{document}